\documentclass[%
aps,
 amsmath,amssymb,
 reprint,%
]{revtex4-1}

\usepackage{graphicx}
\usepackage{dcolumn}
\usepackage{bm}

\begin{document}

\preprint{AIP/123-QED}

\title{Revisiting the fragile-to-strong crossover in metallic glass-forming liquids: \\
application to Cu$_\text{x}$Zr$_\text{x}$Al$_{\text{100-2x}}$ alloy}

\author{Ren\'e Alvarez-Donado}
\email{ralvarez@ifi.unicamp.br}
\affiliation{ 
Instituto de F\'isica Gleb Wataghin, Universidade Estadual de Campinas, UNICAMP, 13083-859 Campinas, S\~ao Paulo, Brazil 
}%
\author{Samuel Cajahuaringa}
\email{samuelif@ifi.unicamp.br}%
\affiliation{ 
	Instituto de F\'isica Gleb Wataghin, Universidade Estadual de Campinas, UNICAMP, 13083-859 Campinas, S\~ao Paulo, Brazil 
}%
\author{Alex Antonelli}
\email{aantone@ifi.unicamp.br}
\affiliation{ 
	Instituto de F\'isica Gleb Wataghin and Center for Computing in Engineering \& Sciences, Universidade Estadual de Campinas, UNICAMP, 13083-859 Campinas, S\~ao Paulo, Brazil 
}%

\date{\today}

\begin{abstract}
The fragile-to-strong crossover seems to be a general feature of metallic glass-forming liquids. Here, we study the behavior of shear viscosity, diffusion coefficient and vibrational density of states for Cu$_\text{x}$Zr$_\text{x}$Al$_{\text{100-2x}}$ alloy through molecular dynamics simulations.  The results reveal that the fragile-to-strong temperature (T$_\text{fs}$) and the glass transition temperature (T$_\text{g}$) increase as the aluminum content becomes larger. The inverse of the diffusion coefficient as a function of temperature exhibits a dynamical crossover in the vicinity of T$_\text{g}$, at a much lower temperature than that predicted by nearly all previous studies. At the temperature in which the dynamical crossover occurs determined by the inverse of the diffusion coefficient, we found an excess of vibrational states at low frequencies, resembling a pronounced peak in the reduced vibrational density of states characteristic of a strong liquid. Finally, the behavior of the shear viscosity as a function of reduced temperature (T$_\text{g}$/T) also shows that, besides the fragile-to-strong crossover nearby T$_\text{g}$, another dynamical crossover is present near the onset of the supercooled regime.  
\end{abstract}

\maketitle

\section{\label{sec:level1}Introduction}
\vspace*{-0.25cm}
Almost sixty years after having synthesized the first metallic glass\cite{Klement1960}, the nature of such materials, as well as other glasses, remains an unsolved problem in condensed matter physics and materials science \cite{Anderson1995,Debenedetti2001}. In particular, the temperature at which a liquid becomes a glass, the glass transition T$_\text{g}$, is strongly dependent on the chemical composition and the cooling rate used to obtain the glass \cite{Naumis,Naumis2006}. Therefore, it is not possible to set a single value for T$_\text{g}$, and different criteria are used to define when a liquid becomes a glass. A common criterion is to choose T$_\text{g}$ as the value at which the shear viscosity of the liquid attains a value of $10^{13}$ poise.\\
\indent Angell\cite{Angell1991} proposed a classification of glasses according to the thermal behavior of the shear viscosity, thus defining the concept of liquid fragility. Glass-forming liquids are called strong when their shear viscosity is reasonably well described by a single Arrhenius equation, presenting a low fragility index. On the other hand, liquids that cannot be described by a single Arrhenius equation are called fragile, and present a high fragility index compared to strong liquids \cite{Debenedetti2001}. For most non-metallic glass-forming liquids a single fragility index describes the full range of temperatures. Nevertheless, some liquids in the supercooled regime, such as water \cite{Ito1999,Chen2006}, silica \cite{Saika-Voivod2001,Saika-Voivod2004}, BeF$_2$ \cite{Hemmati2001}, specific compositions of yttria-alumina \cite{Wilding2002}, and triphenyl-phosphite \cite{J.Wiedersich1997}, do not fit into either of the two categories. Since they present two fragility indexes, for low and high temperatures, such liquids exhibit a dynamical crossover, starting as a fragile liquid and transforming into a strong liquid in the vicinity of T$_\text{g}$. On the other hand, this fragile-to-strong crossover (FSC) seems to be a general feature of MGFLs \cite{Zhang2010,Zhou2015}.\\
\indent In order to understand the reason for this crossover in the supercooled regime several theories have been proposed. A study explaining the glass transition through elastic waves in liquids \cite{Trachenko2009} predicts the FSC in the supercooled regime in the vicinity of T$_\text{g}$ as a consequence of local relaxation events in all system leading to a temperature-independent activation energy. Another attempt at describing FSC is based on the extended mode coupling theory (MCT) \cite{Chong2009}. In such theory, the diffusion coefficient and the relaxation time change their behavior from non-Arrhenius to Arrhenius  at a temperature value approximate to the critical temperature T$_\text{c}$ of the original version of MCT. Computational and experimental works have been carried out in parallel to uncover the nature of FSC. Chen et al \cite{Chen2006}. found a FSC in supercooled water by studying the thermal behavior of the inverse of the self-diffusion coefficient and the average of translational relaxation time. Zhou \cite{Zhou2015} et al. proposed the approximate relation T$_\text{fs}$ $\approx$ 1.36T$_\text{g}$ , based on experimental data for 98 glass-forming liquids. FSC was found experimentally in Zr-based \cite{Zhang2011,Zhou2015} and more recently in Fe-based \cite{Bochtler2017} MGFLs. Moreover, in spite of various theoretical and experimental approaches already utilized in order to understand the reason behind the FSC, its origin remains unclear.\\
\indent Currently, two dynamical crossovers emerge when MGFLs are quenched to form a glass. The first happens near the onset of the supercooled regime and is associated with a transition from an uncorrelated liquid dynamics to a state of increasing cooperation in various regions \cite{Phillips,Langer2007}. During this transition, the transport properties deviate from Arrhenius to non-Arrhenius behavior. The second crossover is the fragile-to-strong one, which occurs in the vicinity of T$_\text{g}$. In this work, we present evidence from computer simulations of FSC for three Cu$_{\text{x}}$Zr$_{\text{x}}$Al$_{100-2\text{x}}$ MGFLs. We choose this type of alloys because it is an ideal model, widely used to explore the properties of metallic glasses \cite{Zhang2011,Zhang2012,Lad2012,Puosi2018,Zhou2015,Li2018,Jaiswal2015,Foroughi2016}. It is known that a small percentage of aluminum increases the fragility index and the glass transition temperature, allowing the exploration of changes in dynamical properties using molecular dynamics simulations \cite{Cheng2008}. The FSC in Cu-Zr-Al alloys has been verified in experimental studies and suggested by simulation analysis \cite{Lad2012,Zhou2015,Zhang2012b}. Notwithstanding, the previous experimental studies are based on values of viscosity obtained at high temperatures when the first dynamical crossover has not happened yet, and, therefore, the value of T$_{\text{fs}}$ could be overestimated. To the best of our knowledge, we are not aware of any computational study that confirms the existence of the FSC in Cu-Zr-Al MGFLs by showing directly the change in the behavior from non-Arrhenius to Arrhenius in either the diffusion coefficient or the shear viscosity. In view of that, we conducted further investigations into the evidence of FSC in metallic glasses through computer simulations. For this purpose, we analyzed the temperature dependence of diffusion coefficient, vibrational density of states and shear viscosity of CuZr(Al) MGFLs using an EAM potential developed by Cheng et al. \cite{Cheng2009}, which adequately describes the properties of Cu-Zr-Al interatomic forces.\\
\indent The remainder of the paper is organized as follows: section II describes the methods used in this study for the calculation of the thermodynamic and dynamic properties of the alloys; results and discussion are presented in section III; and, finally, section IV conveys our conclusion.

\vspace*{-0.65cm}

\section{\label{sec:level2}Methods}

\vspace*{-0.25cm}

\subsection{\label{A} Simulation setup }

\vspace*{-0.25cm}

The computer simulations were carried out employing the widely used molecular dynamics (MD) open code LAMMPS \cite{Plimpton1995}. At first, we started with a binary alloy Cu-Zr, constituted by 4000 atoms (2000 Cu atoms and 2000 Zr atoms) in a cubic box subject to periodic boundary conditions in all three dimensions. An embedded-atom method (EAM) potential was used to describe the interatomic interactions \cite{Cheng2009,Daw1984}. Once the liquid at 2000 K reached the equilibrium, the alloy was quenched to 300 K at a rate of 100 K/ns. The quench was performed in the isobaric-isothermal (NPT) ensemble at zero external pressure. The Nos\'e-Hoover equations were integrated using a time step of 2 fs, with the pressure and temperature damping parameters of 200 fs and 5 ps, respectively. During the quenching process, several configurations were saved in order to obtain the dynamic and thermodynamic properties, which were determined through equilibrium simulations. All correlation functions were calculated in the canonical (NVT) ensemble.\\
\indent We repeated the same procedure for two ternary alloys ($x = 49, 46$) and analyzed the behavior of the viscosity, self-diffusion coefficient and vibrational density of states with the aluminum percentage increase.
\vspace*{-0.5cm}

\subsection{\label{B} Glass transition}
\begin{figure}
\centering
\includegraphics[width=85mm]{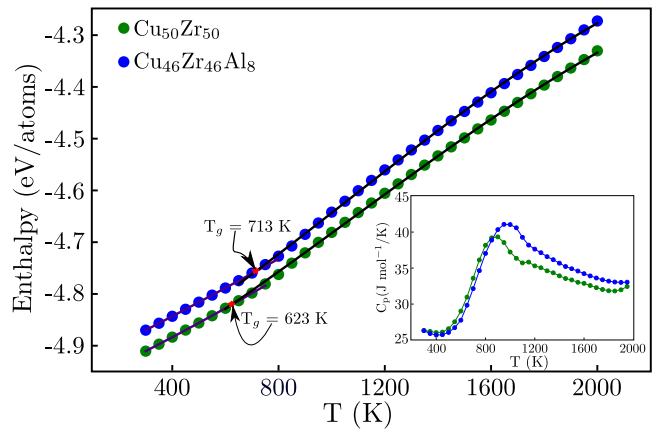}
\caption{ Enthalpy per atom of binary Cu$_{50}$Zr$_{50}$ and ternary Cu$_{46}$Zr$_{46}$Al$_{8}$ alloys as a function of temperature extracted from MD equilibrium simulations. The error bars are of the size of the symbols. Two fitting functions were used to obtain the glass transition temperature (see text). The inset shows the isobaric heat capacity C$_p$ behavior for the same alloys.}
\label{glass}
\end{figure}

\vspace*{-0.25cm}
As the quench was being carried out, we saved atomic configurations every 50 K in order to obtain thermodynamic information during the cooling process. These configurations were used as input in MD equilibrium simulations. After thermalization, thermodynamic properties were determined from MD equilibrium simulations comprising $10^6$ time steps. The behavior of the enthalpy and isobaric heat capacity as a function of temperature for binary Cu$_{50}$Zr$_{50}$ and ternary Cu$_{46}$Zr$_{46}$Al$_{8}$ are presented in Fig.~\ref{glass}.
 The isobaric heat capacity C$_p$ defined through the derivative of the enthalpy:
\begin{equation}
	\text{C}_{p} = \left(\frac{\partial H}{\partial T}  \right)_{p},
    \label{heat}
\end{equation}
was used to estimate the glass transition temperature, T$_\text{g}$. For solid and liquid states, C$_p$ is expressed by the equations:
\begin{equation}
\text{C}_p = 3k_B + cT + dT^2
\label{solid}
\end{equation}\\
and
\begin{equation}
\text{C}_p = 3k_B + aT^2 + \frac{b}{T^2},
\label{liquid}
\end{equation}\\
respectively. Eq.~$(\ref{solid})$ is the solid correction to the Dulong-Petit law and Eq.~$(\ref{liquid})$ is the Kubaschewski relation to describe the C$_\text{p}$ of liquids \cite{Kubaschewski1993}. In order to estimate the value of T$_\text{g}$, we integrate Eqs.~$(\ref{solid})$ and $(\ref{liquid})$ to obtain the enthalpy of the two states. The magenta line in Fig.~\ref{glass} is obtained by fitting the results for the liquid state to the equation for the enthalpy of the liquid, in the interval of 1400 - 900 K, while the orange line is obtained by fitting the data for the solid state to the equation for the enthalpy of the solid, in the range of 500 - 300 K. The estimated values for T$_\text{g}$ are approximately 623 K for the binary alloy and 713 K for the ternary Cu$_{46}$Zr$_{46}$Al$_{8}$, such values were obtained by intersecting the two fitting curves. These results agree with previous experimental and theoretical estimates for T$_\text{g}$ of CuZr(Al) alloys \cite{Zhou2015,Zhang2012a}. A peak in C$_\text{p}$ (inset Fig.~\ref{glass}) is also an evidence that glass transition occurs in that range of temperatures.
\indent For the ternary Cu$_{49}$Zr$_{49}$Al$_{2}$ alloy, a similar procedure was employed to obtain T$_\text{g} \approx $ 632 K; these results are not shown in detail to avoid repetition. 

\indent In Table I the values of T$_\text{g}$ we have determined for each alloy are compared with those obtained experimentally in Ref [15].
\begin{table}[h!]
	\caption{\label{tab:table1}Glass transition temperature T$_\text{g}$ obtained in this work, from experiments, and from other computational studies for the three Cu$_\text{x}$Zr$_\text{x}$Al$_{100-2\text{x}}$ alloys ($x = 50, 49, 46)$ }
	\begin{ruledtabular}
		\begin{tabular}{cccc}
			Alloy & T$_g$ (K)\footnote{This work}  & T$_g$ (K) \footnote{Ref. [15].} \\
			\hline
			Cu$_{50}$Zr$_{50}$ & 623 & 664 \\
			Cu$_{49}$Zr$_{49}$Al$_{2}$ & 632 & 674 \\
			Cu$_{46}$Zr$_{46}$Al$_{8}$ & 713 & 701 \\
		\end{tabular}
	\end{ruledtabular}
\end{table}

\begin{figure*}
\includegraphics[width=17cm]{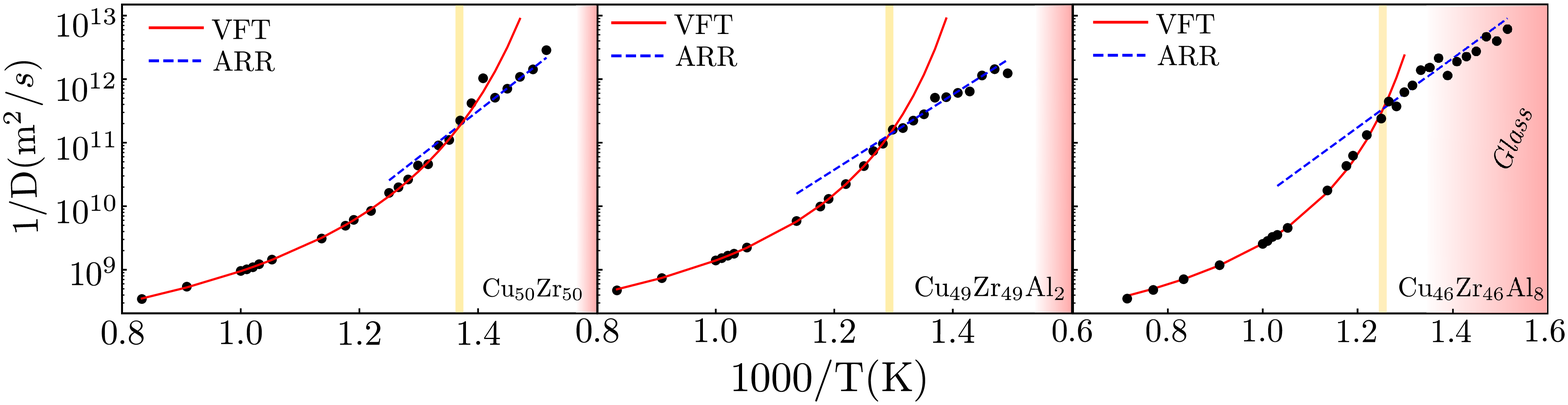}
\caption{\label{diffusion} Temperature dependence of the inverse diffusion coefficient of the three alloys: (a) Cu$_{50}$Zr$_{50}$, (b) Cu$_{49}$Zr$_{49}$Al$_{2}$ and (c) Cu$_{46}$Zr$_{46}$Al$_{8}$. The red line indicates the fitting to a VFT equation and blue dashed is an Arrhenius fit. Yellow shaded region marks the fragile-to-strong crossover.}
\end{figure*}
\vspace*{-0.5cm}
\subsection{Dynamic properties}
\vspace*{-0.25cm}
In order to evidence the FSC, we studied three dynamical properties: diffusion coefficient, vibrational density of states and shear viscosity of the Cu$_{\text{x}}$Zr$_{\text{x}}$Al$_{100-2\text{x}}$ alloys.\\
\indent First, the diffusion coefficient was computed using molecular dynamics simulations through the mean squared displacement (MSD) of each atom
\begin{equation}
\langle r^2(t) \rangle = \frac{1}{N} \left\langle \sum_{l =1}^{N} \left|r_l(t) - r_l(0) \right|^2\right\rangle,
\end{equation}
where $r_l(t)$ is the position vector of the $l^{th}$ particle at time $t$. For long times, the particles are in the diffusive regime, and the MSD is linearly dependent on time. Consequently, the diffusion can be calculated by the equation:
\begin{equation}
D = \lim_{t \rightarrow \infty} \frac{1}{6t}\langle \left|r_i(t) - r_i(0) \right|^2 \rangle.
\label{diff}
\end{equation}
Based on previous studies, an excess in the vibrational density of states (VDOS) of low frequencies with respect to the Debye model is an evidence of a non-Arrhenius-to-Arrhenius behavior crossover in supercooled liquids \cite{Cajahuaringa2013,Jakse2008}. The VDOS is determined using the Fourier transform of the normalized velocity autocorrelation function
\begin{equation}
\text{g}(\omega) = \frac{1}{N} \sum_{l = 1}^{N} \int_{-\infty}^{\infty} \frac{\langle \vec{v}_l(t) \cdot \vec{v}_l(0) \rangle}{\langle \vec{v}_l(0) \cdot \vec{v}_l(0) \rangle}e^{i\omega t}dt,
\label{fonones}
\end{equation}
where $\vec{v}_l$ is the velocity vector of the $l^{th}$ particle at time $t$.\\
At last, the shear viscosity was calculated using a Green-Kubo relation \cite{Hansen,Allen}:
\begin{equation}
\eta = \frac{\beta V}{N} \int_0^{\infty} \langle P_{\alpha\gamma}(0) P_{\alpha\gamma}(t) \rangle dt,
\label{viscosity}
\end{equation}
where $\beta$ is the reciprocal of the absolute temperature times the Boltzmann constant $(k_BT)^{-1}$, $V$ is the volume of the system and $P_{\alpha\gamma}$ is an off diagonal component, $P_{xy}$, $P_{xz}$ and $P_{yz}$, of the stress tensor. Following Alf\`e et. al\cite{Alfe1998}, two more components should be included to calculate the viscosity in Eq. $(\ref{viscosity})$
\begin{equation*}
\frac{1}{2} (P_{xx} - P_{yy})\quad \quad \text{and} \quad \quad \frac{1}{2} (P_{yy} - P_{zz}),   
\end{equation*}
hence, five components were computed in order to obtain the shear viscosity.\\
\indent As we will show later, the crossover temperature T$_{\text{fs}}$ is grater when the Al percentage is increased, and this fact allows us to explore the shear viscosity behavior in the ternary Cu$_{46}$Zr$_{46}$Al$_{8}$ for the values in which the FSC is evidenced. We were not able to observe the crossover from the shear viscosity results of binary Cu$_{50}$Zr$_{50}$ and ternary Cu$_{49}$Zr$_{49}$Al$_{2}$  because the off diagonal stress tensor components do not completely decorrelate even when the sampling time reached 5ns and larger simulation sampling times would exceed our MD capabilities (see Sec III-C).\\
\vspace*{-0.20cm}
\section{Results and discussion}
\vspace*{-0.25cm}
\subsection{Diffusion coefficient}
\vspace*{-0.25cm}
Fig.~\ref{diffusion} shows the inverse of the diffusion coefficient (ID) as a function of the inverse temperature for the three alloys, calculated using Eq. $(\ref{diff})$. We observe a change in the behavior of ID at temperatures near the glass transition for the three alloys. Moreover, this change is shifted toward higher temperatures when aluminum concentration increases. This is due to the properties of MGFLs that strongly depend on composition \cite{Zhang2010,Zhang2011}. The modification in the ID behavior can be associated with the existence of FSC at a certain temperature T$_\text{fs}$, where the crossover takes place. In order to find T$_\text{fs}$ for the alloys, we fitted the data for high temperatures to the  Vogel-Fulcher-Tamman (VFT) equation \cite{VOGEL1921,Fulcher1925,Tammann1926},
\begin{equation}
D = D_0 \exp\left(-\frac{A}{T-T_0}\right),
\label{VFT}
\end{equation}
where $A$ is a parameter that indicates by how much the diffusion diverges from the Arrhenius behavior when the temperature decreases, $T_0$ is the temperature at which the ID diverges and the factor $D_0$ is the diffusion coefficient extrapolated to infinite temperature. Previous studies of CuZr(Al) alloys reveal a dynamical decoupling happening at a temperature T$_\text{A}\approx$ 1200 K \cite{Jaiswal2015,Puosi2018}. This decoupling is associated with the beginning of the dynamical heterogeneities. Above T$_\text{A}$, the liquid alloy presents a non-Arrhenius behavior, therefore for our purposes we fitted our data to a VFT equation for temperatures below T$_\text{A}$, for each alloy.\\
\indent The solid red lines in Fig.~\ref{diffusion} are obtained by the fitting the data to a VFT equation. For low temperatures, close to T$_\text{g}$ (red shaded region in Fig.~\ref{diffusion}), the values for ID are not well fitted by Eq. \ref{VFT}. In this region, it is more appropriate to fit to an Arrhenius equation as follows:
\begin{equation}
D = D_A \exp\left(-\frac{E_A}{k_BT}\right),
\label{Arrh}
\end{equation}
where $E_A$ is the constant activation energy characteristic of a strong liquid \cite{Angell1991,Debenedetti2001}. The dashed blue lines in Fig.~\ref{diffusion} represent the fit to an Arrhenius equation. At the crossover temperature, $T_\text{fs}$, the contributions from the fragile and the strong regions to the whole dynamics are the same, i.e., the temperature at which the values for the diffusion coefficient  calculated from Eqs.~$(\ref{VFT})$ and $(\ref{Arrh})$ are equal. For the binary Cu$_{50}$Zr$_{50}$ the FSC happens at 730 K. However for the ternary alloys Cu$_{49}$Zr$_{49}$Al$_{2}$ and Cu$_{46}$Zr$_{46}$Al$_{8}$ the FSC occurs at 774 K and 797 K, respectively, suggesting a relation between glass transition and FSC as is put forward by Mallamace et.al \cite{Mallamace2010}. Table II shows the values for the fitting parameters and the crossover temperature of the three alloys.\\
\begin{table*}
\caption{\label{table3} Fitting parameters of the VFT and The Arrhenius equations, and fragile-to-strong temperature T$_{fs}$ for the three CuZr(Al) alloys.}
\begin{ruledtabular}
\begin{tabular}{cccccccc}
Alloy & $D_0(m^2/s)\times 10^{-8}$ & A  & T$_0$ (K) & D$_A(m^2/s)\times 10^{-3}$ & E$_A$ (eV) & T$_\text{fs}$ (K) \footnote{This work.}  & T$_\text{fs}$ (K) \footnote{Ref. [15].}\\ \hline
 Cu$_{50}$Zr$_{50}$ & 2.36 & 1333.35 & 571 & 49.23 & 1.44 & 730 & 989 \\
Cu$_{49}$Zr$_{49}$Al$_{2}$ & 1.98 & 1232.81 & 623 & 1.12 & 1.27 & 774 & 1047  \\
Cu$_{46}$Zr$_{46}$Al$_{8}$ & 2.31 & 1272.53 & 650 & 417  & 1.11 & 797 & 1059\\
\end{tabular}
\end{ruledtabular}
\end{table*}
\begin{figure}[h!]
	\centering
	\includegraphics[width=80mm]{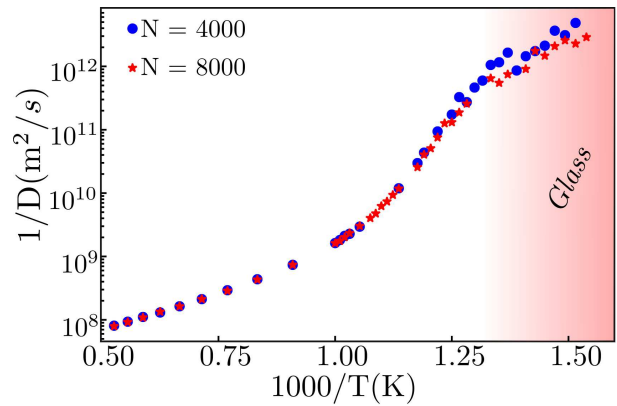}
	\caption{Inverse of the diffusion coefficient as a function of temperature for the ternary Cu$_{46}$Zr$_{46}$Al$_{8}$ metallic alloy for simulation cells N = 4000 atoms and N = 8000 atoms. From the graph one can see that T$_\text{fs}$ remains unaltered.}
	\label{Size}
\end{figure}
The values we have obtained for T$_\text{fs}$ are significantly lower than those previously reported for these alloys \cite{Zhang2011,Zhou2015,Lad2012}. However, in a recent study, Sukhomlinov and M\"user\cite{Sukhomlinov_PRM,sukhomlinov2019anomalous} obtained a value for T$_\text{fs}$ similar to ours, performing molecular dynamics simulations in the Cu$_{29}$Zr$_{60.6}$Al$_{10.4}$ alloy. Since it is very difficult for the experiments to probe directly the molten alloy in the supercooled regime, the experimental estimates of the FSC temperature were determined from models to interpolate the experimental results for viscosity obtained at high temperatures (above the first crossover) and at low temperatures just above the glass transition. \cite{Zhou2015} We believe this is the main reason why the experimental FSC temperature estimates are significantly higher than ours. Regarding the computational studies of the FSC in these alloys, \cite{Zhang2011,Lad2012} their estimates for the FSC temperature are based on indirect evidence of the FSC, instead of a direct observation of the change in behavior from non-Arrhenius to Arrhenius in either the diffusion coefficient or the shear viscosity. This difference in approach can explain the discrepancies between their results and our findings. Following the previous study in Ref \cite{sukhomlinov2019anomalous} about the finite size effects in the FSC for CuZrAl-based metallic glass we explore the size effect in our calculations. In Fig \ref{Size} are shown the behavior of the inverse coefficient diffusion as a function the temperature of the ternary Cu$_{46}$Zr$_{46}$Al$_{8}$ using two supercells, The first one with N = 4000 atoms used in whole our work, and the second one with N = 8000 atoms to show that our results are not changed when the simulation cell increased. As Sukhomlinov and M\"user pointed in their work, size effect decreases with N, this the reason because for T < T$_{fs}$ the fluctuation in the inverse of diffusion coefficient is lesser when N = 4000 than N = 8000.   
\begin{figure}
\centering
\includegraphics[width=80mm]{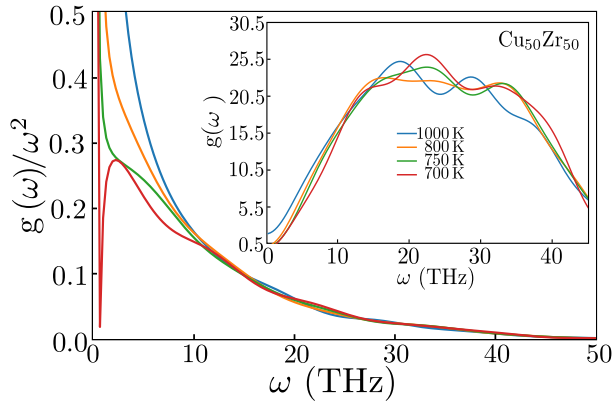}
\caption{ rVDOS and VDOS (inset) of the Cu$_{50}$Zr$_{50}$ alloy as a function of angular frequency for different temperatures. A pronounced boson peak appears for temperatures below 750K, at approximately 1.2 THz }
\label{RVDOS_CuZr}
\end{figure}
\subsection{Vibrational density of states}
\begin{figure}
\centering
\includegraphics[width=80mm]{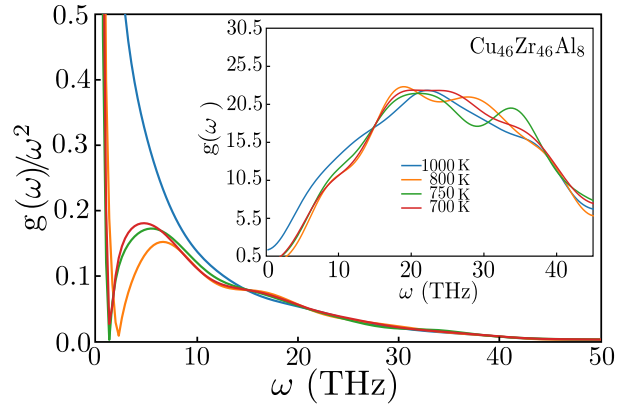}
\caption{ rVDOS and VDOS (inset) of the Cu$_{46}$Zr$_{46}$Al$_8$ alloy as a function of angular frequency for different temperatures. The boson peak that appears below 1000 K shifts towards lower frequencies as the temperature decreases.}
\label{RVDOS_Al8}
\end{figure}

\vspace*{-0.25cm}
Glasses present an excess of vibrational modes in the VDOS compared to Debye law (g($\omega$) $\propto\omega^2)$ at low temperatures\cite{GrigeraT.SMartin-MayorV.ParisiG.2003}. This excess of modes appears as a peak in the reduced VDOS (rVDOS), defined as, g($\omega)/\omega^2$, known as boson peak (BP) \cite{Trachenko2009,Monaco2006,Chumakov2004}. The previous studies about the BP show that the excess in VDOS is more pronounced in strong liquids than in fragile liquids\cite{Sokolov1993}. Therefore, we used as a criterion for the occurrence of the FSC a pronounced BP at low temperatures characteristic of strong liquids.\\
\indent The rVDOS and VDOS for Cu$_{50}$Zr$_{50}$ obtained using Eq. $(\ref{fonones})$ are depicted in Fig. \ref{RVDOS_CuZr}. For temperatures above 1000 K the rVDOS decreases monotonically with increasing $\omega$. Notwithstanding, the deviation from the Debye law appears at 750 K. However, as showed previously in the behavior of ID, the FSC does not occur for this temperature and the BP appears as a weak deviation from the Debye law. Moreover, a pronounced boson peak, appears only at temperatures below 750 K at approximately 1.7 THz, evidencing a change in the behavior of the rVDOS, which could be associated with a FSC in the interval 750-700 K. Those results are consistent with the crossover temperature T$_{\text{fs}}$ obtained from the prior calculation of ID for the same alloy.\\
\indent In  Fig.~\ref{RVDOS_Al8} we show the rVDOS and the VDOS for the ternary alloy Cu$_{46}$Zr$_{46}$Al$_8$. For temperatures above 1000K the liquid does not present any excess modes in the rVDOS, similarly to the case of the binary alloy Cu$_{50}$Zr$_{50}$. At 800 K we observe a pronounced boson peak. Additionally, for 750 and 700 K we see that the peaks are frequency-shifted  in comparison to the binary alloy, happening at approximately 6.8 and 6.2 THz, respectively. In contrast with the binary alloy a pronounced boson peak appears at 800 K and approximately 7.5 THz, evidencing that the FSC is shifted for higher temperatures when the Al percentage is increased.\\
\indent We do not show a similar set of curves for Cu$_{49}$Zr$_{49}$Al$_2$ because the behavior is analogous to the Cu$_{50}$Zr$_{50}$ alloy. Nonetheless, the results are included in supplementary information.\\
\indent In order to provide an additional evidence of the FSC, we analyze the shear viscosity of these CuZr(Al) alloys in the next subsection.
\vspace*{-0.5cm}
\begin{figure}
	\centering
	\includegraphics[width=80mm]{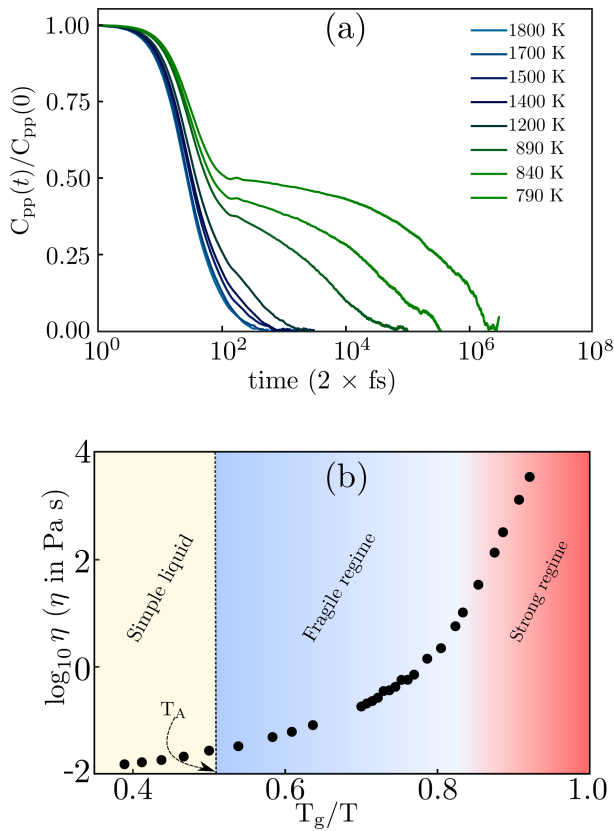}
	\caption{ (a) Averaged autocorrelation function C$_\text{pp}$ as a function of time and (b) logarithm of the shear viscosity as a function of the reduced temperature T$_\text{g}/$T for Cu$_{46}$Zr$_{46}$Al$_{8}$. C$_\text{pp}$ shows a slowing down in the relaxation dynamics for temperatures below 1200 K. Three regimes are well defined in the viscosity behavior, showing the two typical crossovers in supercooled liquids.}
	\label{vis}
\end{figure}
\subsection{Shear viscosity}
\vspace*{-0.25cm}
In order to calculate the shear viscosity using Eq. $(\ref{viscosity})$, the average of the autocorrelation function (C$_\text{pp}$) of the five components of the tension tensor $\langle P_{\alpha\gamma}(t)P_{\alpha\gamma}(0)\rangle$ must be calculated and integrated over all the history of the system. Fig.~\ref{vis} depicts the behavior C$_\text{pp}$ as a function of time for the ternary Cu$_{46}$Zr$_{46}$Al$_{8}$ alloy. For temperatures below 1300 K, C$_\text{pp}$ shows a dramatic slowing down of the relaxation dynamics, this is due to the onset of dynamical heterogeneity that occurs at the characteristic temperature T$_\text{A}$, which is reported to lie in the interval 1000 - 1300 K for the CuZr(Al) metallic alloys \cite{Jaiswal2015,Puosi2018}. For temperatures lower than 790 K, the C$_\text{pp}$ does not decorrelate for $t = 5ns$, presenting a prohibitive computational cost in order to calculate reliable values of viscosity for these temperatures. In the case of binary Cu$_{50}$Zr$_{50}$ and ternary Cu$_{49}$Zr$_{49}$Al$_{2}$, it was not possible calculate C$_\text{pp}$ for temperatures at which the FSC could be observed. The reason why that happens was explained by Cheng et al. \cite{Cheng2008} through the differences between binary and ternary alloys regarding the development of icosahedral clusters as their respective liquids are quenched. By comparing the $\alpha$-relaxation time, which scales with the viscosity, for CuZr and CuZrAl, Cheng et al. concluded that $\tau_{\alpha}$ of the ternary alloy is at least one order of magnitude higher than that of the binary alloy. That is the reason why in our case we were only able to observe the FSC for the ternary alloy Cu$_{46}$Zr$_{46}$Al$_{8}$, since the T$_{\text{fs}}$ is shifted towards higher temperatures when Al percentage increases.\\
\begin{figure}
	\centering
	\includegraphics[width=80mm]{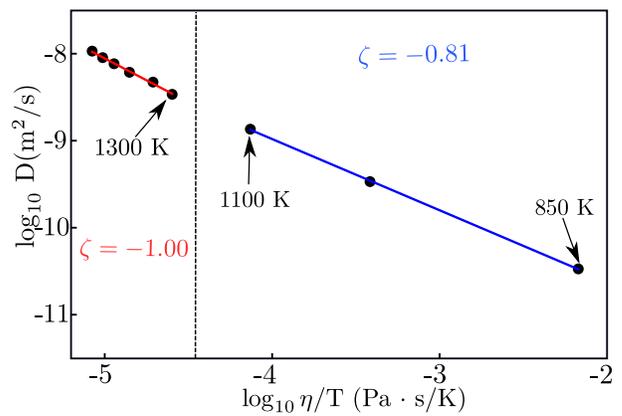}
	\caption{The SE relation for Cu$_{46}$Zr$_{46}$Al$_{8}$. For high temperatures, in the interval 1800 - 1300 K, the slope is $\zeta = -1$. However, for temperatures below T$_A$ = 1200 K, the SE relation deviates from the ideal to a fractional form with $\zeta = -0.81$.}
	\label{stokes}
\end{figure}
\indent The behavior of the logarithm of the shear viscosity as a function of reduced temperature (T/T$_\text{g}$) is displayed in Fig \ref{vis}. Three regimes can be identified during the cooling process. At first, for high temperatures, the alloy presents an Arrhenius behavior when the temperature decreases until approximately 1200 K. This first crossover from Arrhenius to non-Arrhenius behavior is associated with the beginning of the dynamical heterogeneity at specific temperature T$_{\text{A}}$, i.e., the mobilities differ by orders of magnitude in regions few nanometers away. For the case of MGFLs based in CuZrAl, previous studies have shown that this value occurs for a temperature close to 1300 K\cite{Puosi2018,Zhang2015}. Our calculations lead to a value for T$_\text{A}$ of 1200 K, which is in good agreement with previous studies of these alloys \cite{Jaiswal2015}. For temperatures below T$_{\text{A}}$, the shear viscosity develops a non-Arrhenius behavior, which is represented in Fig. \ref{vis} by the blue region. The FSC occurs at the reduced temperature of 0.85 ($\approx 840$ K); again this value for T$_{\text{fs}}$ agrees quite well with those obtained through ID and rVDOS for this alloy. Our estimates for T$_\text{fs}$ are significantly lower than those reported in almost all previous studies of these MGFLs \cite{Zhou2015,Lad2012}.\\
\indent Finally, in Fig.~\ref{stokes} we show the behavior of the logarithm of the diffusion coefficient as a function of corresponding logarithm of the viscosity, both computed at the same temperature, in order to check the Stokes-Einstein relation (SE) for Cu$_{46}$Zr$_{46}$Al$_{8}$ alloy. This relation establishes a connection between the diffusion coefficient and the shear viscosity through the relation $D \sim (\eta/\text{T})^{\zeta}$, which is strictly valid only for $\zeta = -1$. However for the ternary alloy this behavior departs from the ideal one to a fractional relation for temperatures below T$_\text{A}$, with $\zeta = -0.81$. This violation of SE relation from $\zeta = -1.00$ to $\zeta = -0.81$ is associated with a crossover from Arrhenius behavior to non-Arrhenius when the alloys are cooled below T$_\text{A}$.\\ 
\vspace*{-0.7cm}
\section{Conclusions}
\vspace*{-0.3cm}
Three dynamic properties showing the existence of a FSC in Cu$_\text{x}$Zr$_\text{x}$Al$_{\text{100-2x}}$ with x = 50,49,46 were calculated through MD simulations. Two dynamical crossovers are evidenced during the cooling process. The first crossover is associated with the dynamical heterogeneities, and marks onset of the non-Arrhenius behavior before the liquid enters in the supercooled regime. By investigating the ID for temperatures below T$_\text{A}$ for Cu$_{50}$Zr$_{50}$, Cu$_{49}$Zr$_{49}$Al$_{2}$ and Cu$_{46}$Zr$_{46}$Al$_{8}$ we observed two regimes well defined. The first regime exhibits a non-Arrhenius behavior fitted to a VFT equation and a second regime, which displays an Arrhenius behavior. From that we are able to estimate the temperature at which the crossover occurs. Our findings for the FSC temperature are significantly lower than nearly all previous estimates from experiments and computer simulations, notwithstanding, we obtained a value for T$_\text{fs}$ in agreement with recent studies using MD with the same interatomic potential \cite{sukhomlinov2019anomalous,Sukhomlinov_PRM}. The reason for these discrepancies with past works possibly stems from the fact that in these studies the FSC temperature determination relies on indirect evidence of the crossover, instead of the direct observation of the change in behavior from non-Arrhenius to Arrhenius in either the diffusion coefficient or the shear viscosity.\\
\indent The rVDOS reveals a pronounced BP at low temperatures, which can be considered as an evidence of FSC. In the case of the binary Cu$_{50}$Zr$_{50}$ a pronounced BP is only observed for temperatures below 750 K. Notwithstanding the appearing of BP coincides with the value T$_\text{fs}$ obtained through of the ID.\\
\indent We also investigated the brakdown of the SE relation for the Cu$_{46}$Zr$_{46}$Al$_8$ alloy occurs below T$_\text{A}$. Our results agree with previous computational studies of similar alloys \cite{Puosi2018,Zhang2011}.
Finally, we explain the origin of FSC and how the Al presence increases the FSC temperature, T$_{\textbf{fs}}$, within the framework of the potential energy landscape (PEL) theory \cite{stillinger1995topographic,stillinger2015energy}, the PEL of a liquid at high temperatures is composed by basins separated by energy barriers that are uniform throughout the configuration space available for the system. The transition between basis, in this case, corresponds to localized rearrangement of atoms. This process is usually called $\beta$. The uniformity of the energy barriers between basins explains the Arrhenius behavior exhibited by diffusion and viscosity of the liquid at high temperatures. As the liquid is further cooled down to the supercooled regime, the behavior of this properties changes to non-Arrhenius, as dynamical heterogeneities set in. This can be understood as consequence of the change in the topography of the PEL. In this regime, the PEL is formed by metabasins (or craters), which are themselves composed by basins separated by energy barriers that are uniform throughout the metabasin. The energy barriers between metabasins are larger than those separating the basins within the metabasins. In this case, two types of processes can happen: transitions between basins, $\beta$ processes, and transitions between metabasins, usually named $\alpha$. The multiple activation energies of these processes can explain the non-Arrhenius behavior of the liquid. According to the PEL theory, the $\alpha$ processes require the concerted rearrangement of a large number of atoms, which gives a delocalized character to these processes. As the liquid is further cooled, there will be a temperature below which the system will be unable to overcome the energy barriers between metabasins and the $\alpha$ processes will cease to occur. Thus, below that temperature diffusion and viscosity will again display an Arrhenius behavior, since only $\beta$ processes will occur. In other words, this is the temperature at which the fragile-to-strong crossover takes place (T$_\text{fs}$). Therefore, the larger the energy barriers between metabasins the higher T$_\text{fs}$ will be. Since the $\alpha$ process requires the rearrangement of many atoms, our results suggest that addition of a small amount of Al to the binary alloy will increase the energy required for the $\alpha$ transition to take place. Thus providing an insight why the addition of small amounts of Al can increase significantly T$_\text{fs}$.

\vspace*{-0.5cm}
\begin{acknowledgments}
\vspace*{-0.3cm}
We gratefully acknowledge support from the Brazilian agencies CNPq, CAPES, under the Project PROEX-0487, and FAPESP under Grants \#2010/16970-0, \#2013/08293-7, and \#2016/23891-6. The calculations were performed at CCJDR-IFGW-UNICAMP and at CENAPAD-SP in Brazil.

\end{acknowledgments}



%

\end{document}